\documentclass[twoside,fleqn]{ActaStyle}
\usepackage{times}

\usepackage{lscape}
\usepackage{graphicx,epsfig}
\usepackage[tbtags]{amsmath}

\def\authorheading{P. Calabrese et al.}

\def\shorttitle{Critical behavior...}

\begin{document}

\pagerange{1}{6}

\title{%
CRITICAL BEHAVIOR OF VECTOR MODELS WITH CUBIC SYMMETRY. 
}

\author{
P. Calabrese\email{calabres@df.unipi.it}$^*$, A. Pelissetto\email{Andrea.Pelissetto@roma1.infn.it}\,$^\dagger$, E. Vicari\email{vicari@df.unipi.it}\,$^\dagger$,
}
{
$^*$ Scuola Normale Superiore and  I.N.F.N., Piazza dei Cavalieri 7,
 I-56126 Pisa, Italy.\\
$^\dagger$ Dipartimento di Fisica dell'Universit\`a di Roma I\\
$^\ddagger$ Dipartimento di Fisica dell'Universit\`a 
and I.N.F.N., 
Via Buonarroti 2, I-56127 Pisa, Italy.
}

\day{April 5, 2002}

\abstract{%
We report on some  results concerning the effects of cubic 
anisotropy and quenched uncorrelated impurities on multicomponent spin models. 
The analysis of the six-loop three-dimensional series provides an accurate
description of the renormalization-group flow.
}


\section{Cubic-symmetric models} \label{sec1}

The magnetic interactions in crystalline solids with cubic symmetry
like iron or nickel are usually modeled using the
O(3)-symmetric Heisenberg Hamiltonian. However, this is a
simplified model, since other interactions are present.
Among them, the magnetic anisotropy that is induced by the lattice
structure is particularly relevant
experimentally. In cubic-symmetric
lattices it gives rise to additional single-ion contributions, the
simplest one being $\sum_i \vec{s}^{\ 4}_i$.
These terms are usually not considered
when the critical behavior of cubic magnets is discussed.
However, this is strictly justified only
if these nonrotationally invariant interactions, that have the
reduced symmetry of the lattice, are irrelevant in the renormalization-group
(RG) sense.

This question has been extensively investigated during the past decades
\cite{Aharony-76,PV-r}. In the field-theoretical context, one considers the 
$\phi^4$ Hamiltonian and adds all cubic-invariant interactions 
that may be potentially relevant. There are two possible terms:
a cubic hopping term $\sum_{\mu} (\partial_\mu\phi_\mu)^2$
and a cubic-symmetric quartic interaction term $\sum_{\mu} \phi_\mu^4$.
The first term was shown to be irrelevant, 
although it induces slowly-decaying crossover effects \cite{Aharony-76}.
In order to study the second one, one considers
a $\phi^4$ theory with two quartic couplings \cite{Aharony-76}:
\begin{equation}
{\cal H}_c = \int d^d x 
\left\{ \frac{1}{2}(\partial_\mu \phi(x))^2 + \frac{1}{2} r \phi(x)^2 + 
\frac{1}{4!} v_0 \left[ \phi(x)^2\right]^2 +
\frac{1}{4!} w_0 \sum_{i=1}^M \phi_i(x)^4 \right\},
\label{Hphi4cubic}
\end{equation}
where $\phi$ is an $M$-vector field ($M=3$ for magnets).
The fixed-point (FP) structure of the model (\ref{Hphi4cubic}) 
has been investigated 
extensively and there is a general consensus that a critical 
value $M_c$ exists such that, for 
$M < M_c$, the stable FP is the O($M$)-symmetric one, 
while for $M>M_c$ criticality is controlled by a new point with cubic 
symmetry, which is the FP for all RG trajectories starting with 
$w > 0$.
The debated issue is the value of $M_c$. While old studies 
indicated $3 < M_c < 4$, recent field-theoretical works
find $M_c < 3$ \cite{PV-r}; more precisely, 
$M_c \approx 2.9$~\cite{KT-95,Varnashev-00,CPV-00,FHY-01}.
This result has several important implications for magnets for which $M=3$. 
If the system tends to
magnetize along the cubic axes---this corresponds to a negative
coupling $w$---then the system undergoes a first-order phase
transition, since it is not in the basin of attraction of the cubic FP and 
therefore the RG flow runs away to infinity. 
Instead, magnets in which 
the cubic interaction favors the alignment of
the spins along the diagonals of the cube, so that $w_0 > 0$,
have a critical behavior with a new set of critical exponents and do not 
show Goldstone excitations even at the critical point. 
However, distinguishing the cubic and
Heisenberg universality classes is expected to be a hard task in practice.
Indeed, the critical exponents differ very little \cite{CHPRV-02,PV-r}:
the cubic exponents are \cite{noi}
$\nu_{c}=0.7109(6)$, $\eta_c=0.0374(5)$, and $\gamma_c=1.3955(12)$,
and
\begin{equation}
\eta_{c} - \eta_H =-0.0001(1) , \quad
\nu_{c} - \nu_H = -0.0003(3) , \quad
\gamma_{c} - \gamma_H = -0.0005(7) ,\label{diffexp}
\end{equation}
These differences are much smaller than the typical experimental errors, 
so distinguishing cubic and O(3) universality class should be very hard.

The results for the cubic model (\ref{Hphi4cubic})
have implications 
for other models. First, we should mention the antiferromagnetic three- and
four-state Potts models. In fact,
as argued in \cite{BGJ-80,Itakura-99}, the critical 
behavior of these models at the high-temperature transition 
should be described by the cubic Hamiltonian 
${\cal H}_c$ with $M=2, 3$ and $w_0<0$.
The results presented above allow us to make the following predictions.
If the three-state model has a critical transition, 
it should belong to the XY universality class. On the other hand, 
the four-state model is expected to show a first-order transition.

We can also use the above-presented results to discuss the nature 
of the bicritical point in models with symmetry O$(N_1)\oplus {\rm O}(N_2)$. 
Indeed, they allow to  exclude that the bicritical point has enlarged symmetry 
O$(N_1 + N_2)$ if $N_1 + N_2 > 2$ \cite{noi2}. 
This result has important implications for the SO(5) theory of 
superconductivity  \cite{Zhang-97}. In the SO(5) theory
\cite{Zhang-97}, one considers a model with symmetry O(3)$\oplus$U(1) =
O(3)$\oplus$O(2) with two order parameters: one is related to the
antiferromagnetic order, the other one is associated with
$d$-wave superconductivity. The main issue is whether the 
SO(5) symmetry can be realized at a bicritical
point where two critical lines, with symmetry O(3) and O(2) respectively,
meet. In RG terms, this can generally occur if
the O(5) FP has only {\em two} relevant
O(3)$\oplus$ O(2)-symmetric perturbations. 
But, when $N\geq 3$,
the instability of the O($N$) fixed point with respect to 
the cubic perturbation shows that at least
another relevant perturbation exists.
The stable FP is expected to be the tetracritical
decoupled FP which can be shown to be stable by 
nonperturbative arguments \cite{Aharony-02}.

Another important class of systems are uniaxial antiferromagnets in a 
magnetic field parallel to the field direction \cite{KNF-76}.
In this case $N_1 = 1$ and $N_2 = 2$. The results presented above show also
that the bicritical O(3)-symmetric fixed point is unstable.
The multicritical behavior should be controlled by the
biconal FP \cite{noi2}, which, however,
is expected to be close to the O(3) FP,
so that critical exponents should be very close to the Heisenberg ones.
Thus, differences should be hardly distinguishable in experiments.

\section{Random impurities and softening: General considerations}
\label{sec2}

\begin{figure}[b]
\begin{minipage}[t]{6.9cm} 
\centerline{\psfig{width=6.9truecm,file=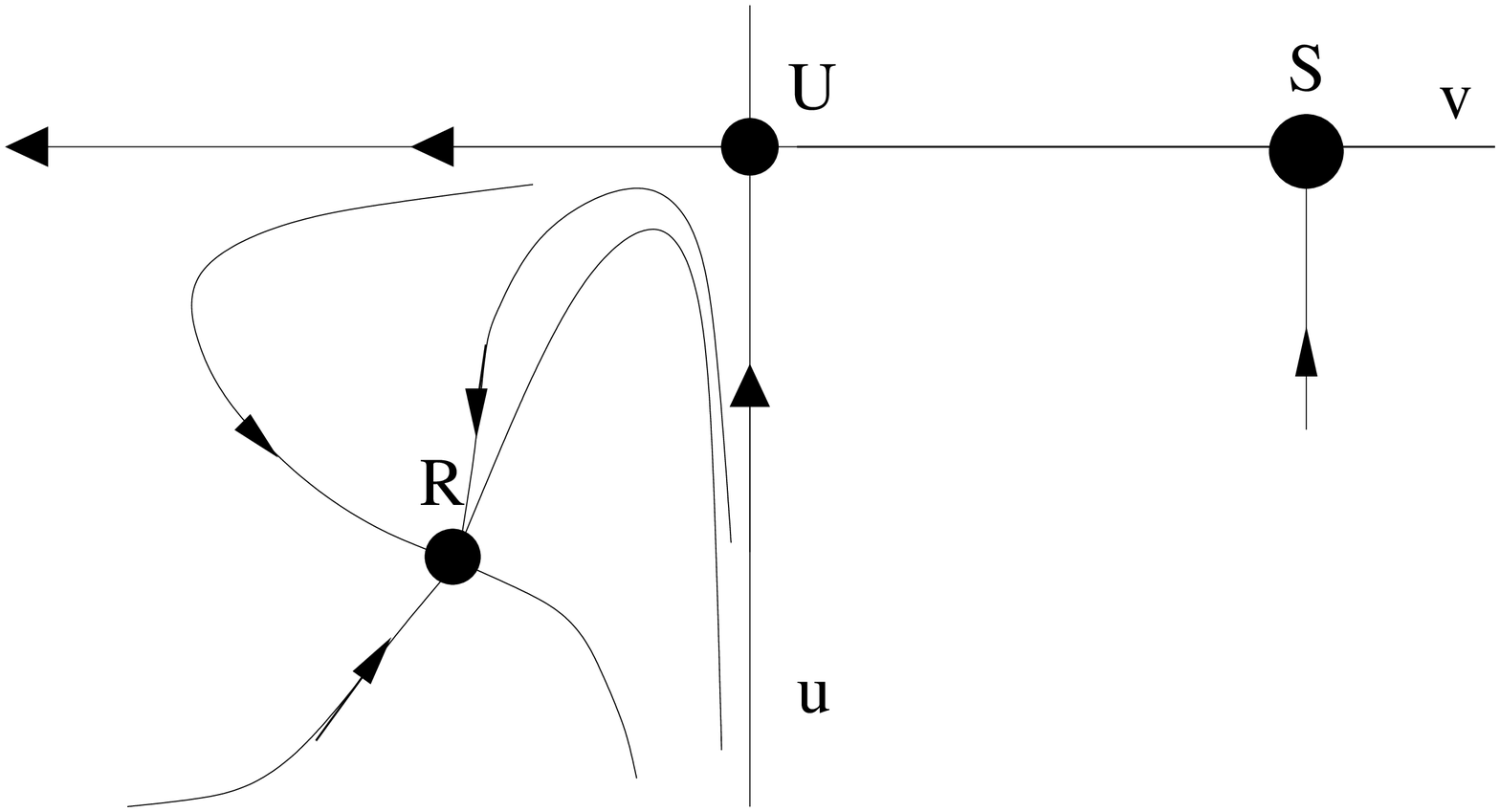}}
\end{minipage} 
 \hspace{2mm} \ 
\begin{minipage}[t]{6.0cm} 
\centerline{\psfig{width=6.0truecm,file=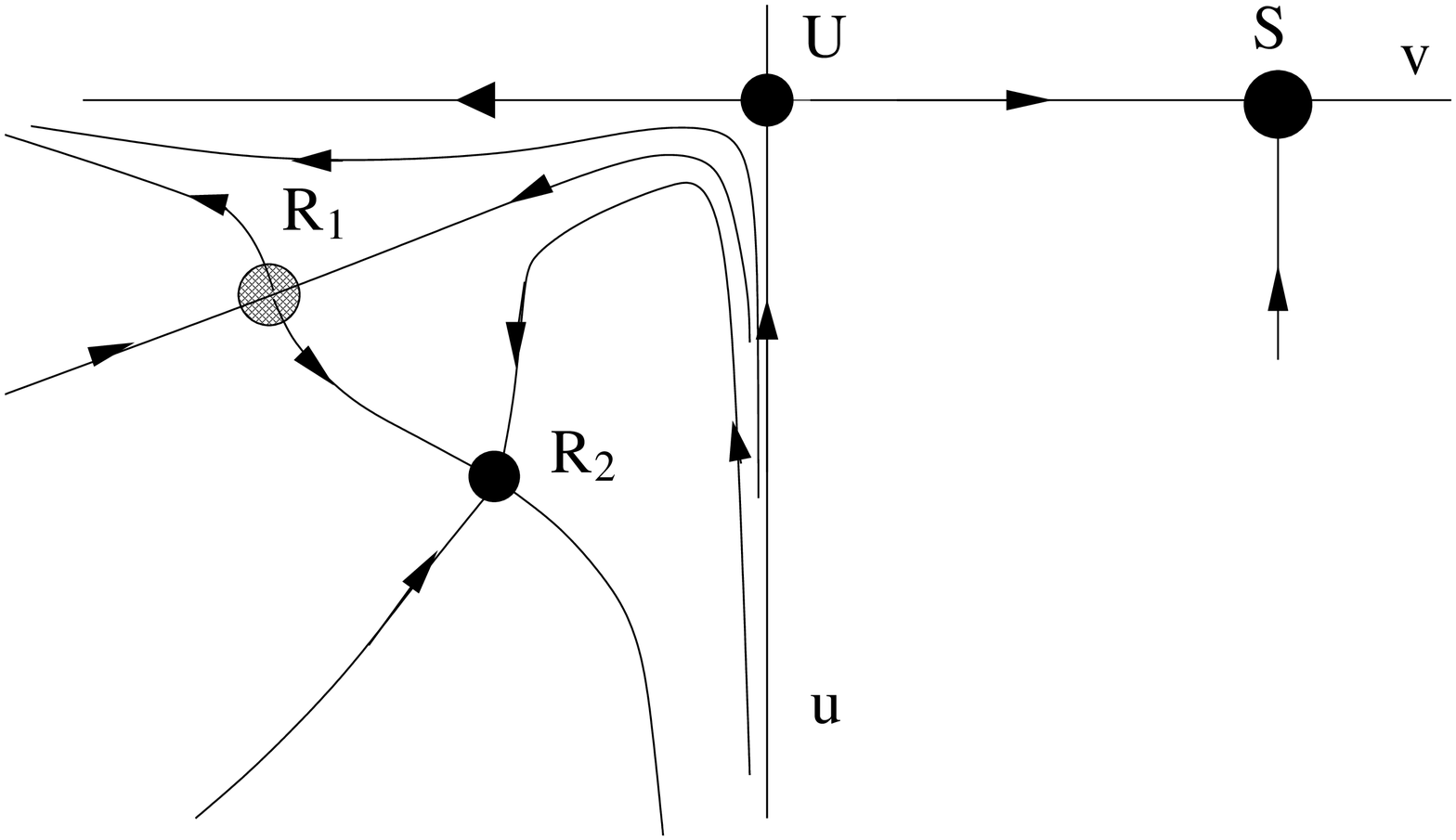}} 
\end{minipage} 
\caption{Softening scenarios for fluctuation-induced first-order transitions.}
\label{fig}
\end{figure}

The critical behavior of systems with quenched disorder is of considerable 
interest. 
Experimentally, dilute systems can be obtained by mixing an 
(anti)-ferromagnetic material with a nonmagnetic one or by considering
a fluid in a porous material, for instance in Vycor.
A practical tool in the study of the effect of randomness on second-order
phase transitions is provided by
the Harris criterion~\cite{Harris-74}. It states that the addition of 
impurities to a system which undergoes a second-order 
phase transition does not change the critical behavior 
if the specific-heat critical exponent $\alpha_{\rm p}$ of the pure 
system is negative. If $\alpha_{\rm p}$ is positive, the transition
is altered. 
Moreover, even if the pure-system FP remains stable, 
disorder may still have physical consequences.
It may change the attraction domain of the pure stable 
FP so that some pure systems undergoing a first-order transition 
in the absence of disorder may show a critical behavior for some 
dilution. Softening of the phase transition may also occur if new 
stable FP's are generated by disorder. Two possible scenarios are 
illustrated in Fig. \ref{fig}. The pure system corresponds to $u=0$ 
and has two FP's: one, labelled $S$, with $v > 0$ is stable, while 
the second one, labelled $U$, with $v=0$ is unstable. Therefore, 
pure systems with $v > 0$ show a critical behavior controlled by $S$,
while systems with $v < 0$ undergo a first-order phase transition.
Then, we introduce randomness in the system, which corresponds to 
considering strictly negative values of $u$. The pure FP is stable against 
disorder and indeed RG trajectories with $v > 0$ still flow towards $S$:
disorder does not change the critical behavior. On the other hand, disorder is
relevant for $v < 0$, if new FP's outside the
basin of attraction of $S$ appear, as illustrated in Fig. \ref{fig}. 
If the scenario on the left occurs, systems corresponding to  $v < 0$ show now
a critical transition that belongs to a new universality class 
controlled by the new random FP. If the scenario on the right 
of Fig. \ref{fig} applies, 
the behavior is more complex.
The transition remains first order for low enough impurity concentration, 
then for a given concentration becomes continuous and in the 
universality class of the unstable (tricritical) FP, and finally, 
for larger concentrations (but still under the percolation threshold) it is in
the attraction domain of the stable FP. 
In Fig. \ref{fig}
 we have assumed that the attraction domain does not change, but 
it possible that the boundary of the basin of attraction is not 
the line $v = 0$. Therefore, it could also be possible that some 
systems with $v > 0$ do not have a critical behavior controlled by $S$ 
in the presence of disorder, or the opposite case, i.e. that 
systems with $v < 0$
have $S$ criticality in the presence of disorder. 
Another exotic possibility was found by Cardy in two dimensions \cite{Cardy-96}.
In this case, the pure stable FP is marginally unstable against disorder, 
but the RG trajectories are closed
paths starting and finishing in the pure FP, so that the critical 
behavior is unchanged. This peculiar FP
occurs only when disorder is  marginally unstable.

The occurence of softening has been studied carefully for the two-dimensional 
case.
It was argued in Ref.~\cite{IW-79}, and later put on a rigorous basis
\cite{AW-89}, that in two dimensions thermal first-order transitions 
become continuous in the presence of quenched disorder coupled to the local 
energy density. Such a conclusion was confirmed by Cardy that showed 
for a very specific model that softening persists
in $2+\varepsilon$ dimensions \cite{Cardy-96}. 
However, in three dimensions
the analysis of \cite{IW-79} shows that
the occurrence of softening may depend on nonuniversal  
features.

\section{Randomly dilute $M$-vector cubic models} \label{sec3}

The critical properties of dilute $M$-vector models
are often described in terms of an 
O$(M)$-symmetric $\phi^4$ Hamiltonian; the presence of
uncorrelated random impurities is taken into account by
coupling a random field with the local energy density.
Using the Harris criterion, one sees that for $M\ge 2$
the pure FP is stable against disorder since $\alpha_p < 0$. 
On the other hand, in the Ising case the specific-heat exponent
is positive and thus disorder is relevant: the random Ising model
(RIM) shows a new type of critical behavior as confirmed 
experimentally and theoretically, see, e.g., 
Refs.~\cite{random,FHY-01,Varnashev-00,PV-r} and references therein.

The Harris criterion also allows to determine the stability 
properties of the stable FP of the cubic-symmetric model. 
Since $\alpha_p < 0$ in all cases, the stable FP is 
unchanged. Of course, as discussed in the previous section, 
this does not exclude the presence of new stable FP's due to
disorder---they indeed do appear in two dimensions
\cite{Cardy-96}---and therefore the softening of the first-order 
transition observed in pure systems with $w < 0$. 

In order to investigate this possibility, we consider the 
Hamiltonian \cite{Aharony-75}
\begin{equation}
{\cal H}_e = \int d^d x 
\left\{ \sum_{i,a}\frac{1}{2} \left[ (\partial_\mu \phi_{a,i})^2 + 
         r \phi_{a,i}^2 \right] + 
\sum_{ij,ab} \frac{1}{4!}
\left( u_0 + v_0 \delta_{ij} +w_0 \delta_{ij}\delta_{ab}\right)
          \phi^2_{a,i} \phi^2_{b,j} 
\right\},
\label{Hphi4}
\end{equation}
where $a,b=1,...M$ and $i,j=1,...N$.
Using the standard replica trick, one may show that dilute cubic-symmetric 
systems are recovered in the limit $N\rightarrow 0$. 
The coupling $u_0$ is negative, being proportional to minus the variance of
the quenched disorder.

In order to see whether new stable FP's with $u < 0$ exist, we shall use
the $\epsilon$ and the fixed-dimension expansion.
First, we analyze the special cases $v=0$ and $w=0$.
For $v=0$ the Hamiltonian (\ref{Hphi4})
describes an $MN$-component model with cubic anisotropy, characterized by 
the presence of two stable FP's \cite{Aharony-76,PV-r}. 
The one for $u>0$, $w=0$ is in the self-avoiding walk 
(SAW) universality class, but it is irrelevant for our problem, since
it is unreachable from the physical region $u<0$.
The other, with $u<0$, $w>0$, belongs to 
the RIM universality class. 
In the case $w=0$, the Hamiltonian (\ref{Hphi4}) describes $N$ coupled
$M$-vector models, and it is also called $MN$ model \cite{Aharony-76}.
Again, the flow is characterized by two stable FP's:
the SAW and the O($M$)-symmetric ones. They are both irrelevant for our 
problem. The SAW FP has $u > 0$, while the O($M$)-symmetric one 
has $u=0$ and it is unstable against $w$ perturbations.
For $M=2$ and generic $N$,
the Hamiltonian (\ref{Hphi4}) is invariant under a general
transformation, cf. Ref.~\cite{Korz-76}.
For $N=0$, it maps the RIM FP into a new RIM FP belonging to the region with 
$u<0$, $v>0$, $w<0$. 

The above-reported considerations show the presence of only 
one (for $M=2$ two) FP that could be possibly stable: the RIM 
FP with $v=0$ and, for $M=2$, the second RIM FP related to the 
previous one by symmetry. Of course, other FP's may have 
$u < 0$, $v\not=0$, $w\not=0$ and thus a more general analysis is 
needed in order to have a complete knowledge of the RG flow.

The RG flow can be investigated
near four dimensions using the perturbative $\epsilon$ expansion. 
The results are reported in \cite{noi}.  
No new FP's (apart the one predicted by 
symmetry \cite{Korz-76} for $M=2$) are found 
in the region of physical interest $u<0$.
Thus, near four dimensions the critical behavior is not changed by 
the addition of random impurities for any $M\geq 2$.
Moreover, there is no  softening of the transition for pure systems
that are outside the attraction domain of the stable FP.

The $\epsilon$-expansion analysis shows that the random FP that is found in 
$2+\epsilon$ dimensions \cite{Cardy-96} eventually disappears as 
the dimension is increased. The interesting question is therefore if, 
for $d=3$, one observes a qualitative behavior analogous to the 
two-dimensional or four-dimensional case. Such a question can only be 
investigated in a strictly three-dimensional scheme. 
For this reason, we computed the RG functions to six loops in the 
fixed-dimension expansion \cite{noi}, and carefully investigated the 
FP structure of the model.

First of all, we checked the stability properties of the already
known FP's.
For all values of $M$ we found, in agreement with the Harris criterion, 
that the pure stable FP remains stable after dilution.
Furthermore, the numerical estimate of the crossover exponent
agrees with the theoretical prediction $\phi_u=\alpha_{\rm p}/\nu_{\rm p}$.

We also studied  the stability of the RIM FP in the plane $v=0$,
computing the $M$-independent crossover exponent $\phi_v$ at the 
RIM FP. Our final estimate $\phi_v=0.04(5)$ suggests that the 
$v$ perturbation is relevant and thus that the RIM FP is 
unstable, although the relatively large error bar does not allow us to 
exclude the opposite case.
This point deserves further investigations, for example using other
resummation methods.
Note that, even if the perturbation is relevant, the RG dimension is very small
and thus one expects very strong crossover effects.

We searched for the presence of new FP's. 
For $M=2$, we  only found the FP predicted by symmetry.
For larger $N$, our analysis did not provide evidence for new FP's in 
the physical region $u <0$.
Therefore, no softening is expected, at least in the region of sufficiently low
impurity concentration where the field-theoretical approach is justified. 

Finally, we mention that the above results can be used to
determine the critical behavior of dilute three- and four-state
antiferromagnetic Potts models, which should be  described by the
dilute cubic model (\ref{Hphi4}) in the two- and three-component
cases respectively and for $w_0<0$ \cite{noi}.
They imply that  the dilute
three-component antiferromagnetic Potts
model presents a continuous transition belonging to the
XY universality class, as in the pure case.
In the four-state case, the weak first order transition
expected in the pure case should not be softened by 
random dilution.


\end{document}